\documentclass[reprint, amsmath, amssymb, aps, showkeys]{revtex4-2}
\usepackage{graphicx}
\usepackage{dcolumn}
\usepackage{bm}
\usepackage{amsmath}
\usepackage{float}
\usepackage{subfigure}
\usepackage{booktabs}
\usepackage{CJKutf8}
\usepackage{hyperref}
\usepackage{tikz,xcolor}

\hypersetup{
  colorlinks=true,
  linkcolor=blue,
  urlcolor=blue,
  citecolor=blue,
  linktoc=all
}
\definecolor{lime}{HTML}{A6CE39}
\DeclareRobustCommand{\orcidicon}{%
	\begin{tikzpicture}
	\draw[lime, fill=lime] (0,0) 
	circle [radius=0.16] 
	node[white] {{\fontfamily{qag}\selectfont \tiny ID}};
	\draw[white, fill=white] (-0.0625,0.095) 
	circle [radius=0.007];
	\end{tikzpicture}
	\hspace{-2mm}
}

\foreach \x in {A, ..., Z}{%
	\expandafter\xdef\csname orcid\x\endcsname{\noexpand\href{https://orcid.org/\csname orcidauthor\x\endcsname}{\noexpand\orcidicon}}
}

\begin{document}
\begin{CJK*}{UTF8}{gbsn}

\title{Effect of kick velocity on gravitational wave detection of binary black holes\\ with space- and ground-based detectors}

\author{Jie Wu (吴洁)\orcidA{}$^{1,2}$ }
\author{Mengfei Sun (孙孟飞)\orcidB{}$^{1,2}$ }
\author{Xianghe Ma (马翔河)\orcidC{}$^{1,2}$ }
\author{\\ Xiaolin Liu (刘骁麟)\orcidD{}$^{4}$}
\author{Jin Li (李瑾)\orcidE{}$^{1,2,3}$ }
\email{cqujinli1983@cqu.edu.cn}
\author{Zhoujian Cao (曹周键)\orcidF{}$^{5,6,7}$ }
\email{zjcao@bnu.edu.cn}

\affiliation{$^{1}$College of Physics, Chongqing University, Chongqing 401331, China}
\affiliation{$^{2}$Department of Physics and Chongqing Key Laboratory for Strongly Coupled Physics, Chongqing University, Chongqing 401331, China}
\affiliation{$^{3}$Institute of  Advanced Interdisciplinary Studies,Chongqing University, Chongqing 401331, China}
\affiliation{$^{4}$Institudo de Física Téorica UAM-CSIC, Universidad Autónoma de Madrid, 28049, Spain}
\affiliation{$^{5}$Department of Astronomy, Beijing Normal University, Beijing 100875, China}
\affiliation{$^{6}$Institute for Frontiers in Astronomy and Astrophysics, Beijing Normal University, Beijing 102206, China}
\affiliation{$^{7}$School of Fundamental Physics and Mathematical Sciences, Hangzhou Institute for Advanced Study, UCAS, Hangzhou 310024, China}

\begin{abstract}
During the coalescence of binary black holes (BBHs), asymmetric gravitational wave (GW) emission imparts a kick velocity to the remnant black hole, affecting observed waveforms and parameter estimation.  
In this study, we investigate the impact of this effect on GW observations using space- and ground-based detectors.
By applying Lorentz transformations, we analyze waveform modifications due to kick velocities.  
For space-based detectors, nearly 50\% of detected signals require corrections, while for ground-based detectors, this fraction is below one-third.
For \texttt{Q3d} population model, space-based detectors could observe kick effects in over 60\% of massive BBH mergers, while in \texttt{pop3} model, this fraction could drop to 3$\sim$4\%. 
Third-generation ground-based detectors may detect kick effects in up to 16\% of stellar-mass BBH mergers.
Our findings highlight the importance of incorporating kick velocity effects into waveform modeling, enhancing GW signal interpretation and our understanding of BBH dynamics and astrophysical implications.
\end{abstract}

\maketitle
\end{CJK*}

\section{Introduction}
The detection of GW150914 not only confirmed the predictions of general relativity but also heralded the dawn of gravitational wave (GW) astronomy~\cite{GW150914}.
Since then, the LIGO-Virgo-KAGRA (LVK) collaboration has detected nearly a hundred GW events, marking significant progress in this field~\cite{IAS1,IAS2,pycbc_group,aLIGO,aVIRGO,KAGRA}. 
To further enhance sensitivity to weaker signals, the development of third-generation detectors such as the Einstein Telescope (ET) is underway~\cite{ET}. 
Additionally, space-based detectors like LISA~\cite{LISA}, Taiji~\cite{Taiji}, and TianQin~\cite{TianQin} are anticipated to begin operations in the 2030s. 
The advancements in ground- and space-based detectors promise to provide a wealth of data and broaden the scope of GW astronomy.

Among the diverse sources of GWs, binary black holes (BBHs) hold particular significance. 
To date, all observed GW events have originated from binary compact object mergers, most of which are BBHs~\cite{GWTC1,GWTC2,GWTC3}. 
During the coalescence process, BBHs lose energy, linear momentum, and angular momentum through gravitational radiation, leading to orbital decay and eventual merger~\cite{kick1,kick_more2,kick_more3}. 
The final state of the remnant black hole (BH) is determined by the initial configuration of the BBH~\cite{kick3,kick_more4,kick_more1}. 
Due to the anisotropic emission of GWs, BBHs experience a net loss of linear momentum, resulting in a recoil or kick velocity that can reach thousands of km/s~\cite{kick4,kick_more5,kick_more6}. 
The first identification of a large kick velocity for an individual GW event (GW200129) has the velocity constraint of $\sim$1500~km/s, at 90\% credibility~\cite{first_kick_event1,first_kick_event2}.
This kick effect is not only a critical aspect of GW physics but also has profound astrophysical implications~\cite{kick2}. 
For instance, in the case of stellar-mass binary black holes (SBBHs), the remnant BH may participate in subsequent mergers~\cite{kick_SBBH1,kick_SBBH2,kick_SBBH3}. 
Furthermore, for massive black hole binaries (MBHBs), the remnant can significantly influence the formation and evolution of its host galaxy~\cite{kick_MBHB1,kick_MBHB2,kick_MBHB3}.

Given the importance of the kick effect, it is essential to develop accurate waveform models that incorporate the complete inspiral-merger-ringdown (IMR) signal. 
Recent studies have made some significant strides in this direction. 
Gerosa~\textit{et al.} employed GW waveform modeling to efficiently extract kick velocities from typical BBH systems~\cite{surrkick}. 
Varma~\textit{et al.} demonstrated that accounting for kick effects is crucial for avoiding systematic biases in data analysis, particularly with third-generation detectors~\cite{kick3}. 
References~\cite{Surrogate1} and \cite{Surrogate2} use a surrogate model to predict the GW waveform and remnant BH characteristics including kick effects.
Mahapatra~\textit{et al.} were the first to quantify the precision with which future detectors can measure the impact of kicks on gravitational waveforms~\cite{test_GR_kick}.
Reference~\cite{kick4} introduced a method analogous to cosmological redshift, rescaling the total mass to calibrate waveform models. 
In Ref.~\cite{paper_Cao}, He~\textit{et al.} derived the explicit Lorentz transformation of the GW tensor, enabling the construction of waveforms for sources moving at arbitrary velocities. 
Despite these advancements, the influence of full kick velocity on observations across different detectors remains underexplored.

Building on these foundational works and our previous studies~\cite{paper1,paper2,paper3}, we present a complete IMR waveform model that incorporates the full kick velocity effect. 
Our study systematically examines the impact of kick velocity on GW detection by considering its temporal evolution and components. 
We calculate the time transformation and Lorentz transformation of the GW tensor across various velocity frames and perform waveform modeling in different parameter spaces. 
Simulations are conducted using ground-based detectors (LVK and ET) for SBBHs and space-based detectors (LISA, Taiji, and TianQin) for MBHBs. 
Additionally, we evaluate the influence of kick velocity on detection by analyzing several typical BBH population models. 
This comprehensive investigation provides a robust framework for determining when and how kick velocity should be accounted for in GW observations.

The structure of this paper is as follows. 
In Sec.~\ref{sec:GW_Signal}, we review the computation of kick velocity and the transformation of GW waveforms for moving sources. 
Section~\ref{sec:Detectors} introduces the detectors used in this study, including their configurations and noise power spectral densities (PSDs). 
In Sec.~\ref{sec:Methodology}, we describe the methodology for calculating the signal-to-noise ratio (SNR), define the parameter spaces for simulations, and introduce several BBH population models. 
Section~\ref{sec:Results} presents our simulation results, highlighting the effects of kick velocity on waveforms across different parameters and the outcomes of observing various BBH population models with different detectors. 
Finally, we summarize our key findings and conclusions in Sec.~\ref{sec:Conclusion}. 
Throughout this paper, we use natural units with $c=G=1$, where $c$ is the speed of light and $G$ is the gravitational constant.

\section{GW Signal}\label{sec:GW_Signal}
\begin{figure}[ht]
    \begin{minipage}{\columnwidth}
        \centering
        \includegraphics[width=0.8\textwidth,
        trim=0 0 0 0,clip]{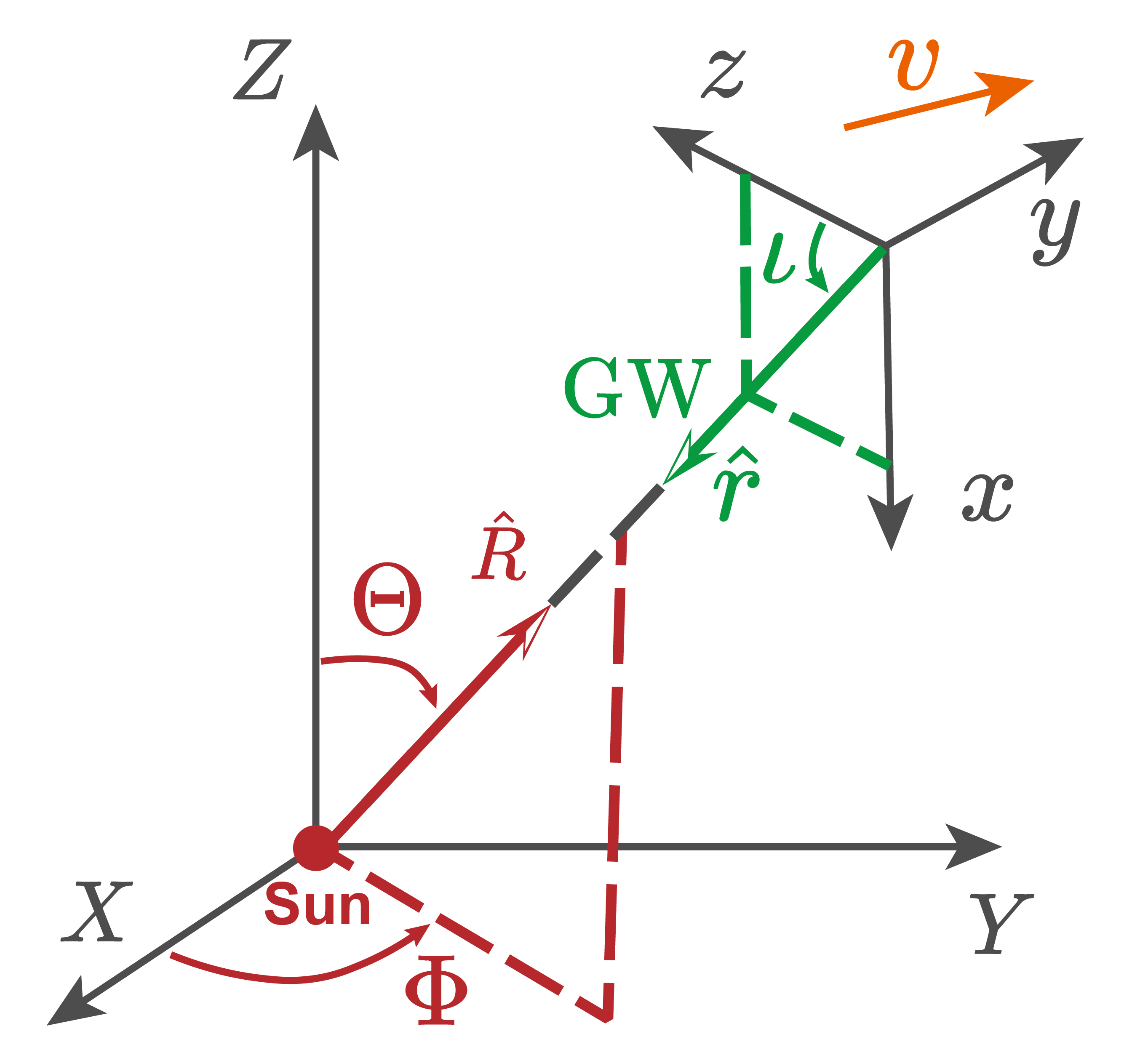}
        \caption{The diagram of the SSB frame ($X,Y,Z$) and source frame ($x,y,z$). We define a coordinate system where the $z$-axis aligns with the orbital angular momentum of the BBH, and the $x$-$z$ plane includes the line connecting the BBH and the detector.}\label{fig:frame}
    \end{minipage}
\end{figure}

In the framework of general relativity, GW exhibits two distinct polarizations. 
These can be mathematically described in tensor form as
\begin{equation}
	h_{ij}=h_+e^+_{ij}+h_\times e^\times_{ij},
\end{equation}
where $h_+$ and $h_\times$ are the GW waveforms of plus and cross modes, $e^+_{ij}$ and $e^\times_{ij}$ are the polarization tensors. 
Utilizing the solar system barycenter (SSB) frame, we establish a right-handed orthonormal basis to construct the GW polarization tensor (see Refs.~\cite{paper1,GWSpace,TDC}).
For GWs in the rest frame, we employ \texttt{SEOBNRE}, which is the latest Effective-One-Body Numerical-Relativity waveform model for Eccentric spin-precessing BBH coalescence~\cite{SEOBNRE1,SEOBNRE2}.
The reference frames are established as depicted in Fig.~\ref{fig:frame}, where $\hat{r}$ denotes the propagation direction of the GW.

In four-dimensional spacetime, GWs exhibit the characteristics of boost weight zero and spin weight 2, which complicates the understanding of Lorentz transformation. 
Moreover, GW represented in the form of a three-dimensional tensor can be calculated and transformed in any coordinate system~\cite{book1,book2}. 
Hence, the Lorentz transformation of GW can be conducted via the three-dimensional Lorentz tensor transformation. 
A comprehensive derivation of this transformation is provided in Ref.~\cite{paper_Cao}. 
The GW waveform of moving source with arbitrary velocity can be constructed as
\begin{equation}\label{eq:Lorentz1}
	\begin{aligned}
		h_{ij}' & =h_{ij}+v^{k}h_{kl}v^{l}\frac{1}{(1-\hat{r}\cdot\vec{v})^{2}}[\hat{r}_{i}\hat{r}_{j} \\
		 & -\frac{\gamma}{1+\gamma}(\hat{r}_{i}v_{j}+v_{i}\hat{r}_{j})+\frac{\gamma^{2}}{(1+\gamma)^{2}}v_{i}v_{j}] \\
		 & +v^{k}h_{kj}\frac{1}{1-\hat{r}\cdot\vec{v}}[\hat{r}_{i}-\frac{\gamma}{1+\gamma}v_{i}] \\
		 & +v^{k}h_{ik}\frac{1}{1-\hat{r}\cdot\vec{v}}[\hat{r}_{j}-\frac{\gamma}{1+\gamma}v_{j}], 
		\end{aligned}
\end{equation}
with
\begin{equation}
	\gamma=\frac{1}{\sqrt{1-v^{2}}},
\end{equation}
where $\vec{v}$ is the velocity of the moving frame relative to the rest frame.
The coordinate time transformation resulting from the Lorentz transformation can be expressed as
\begin{equation}\label{eq:Lorentz2}
	h'_{ij}(t)=h'_{ij}(t'/k),
\end{equation}
with
\begin{equation}
	k=\frac{1}{\gamma(1-\vec{v}\cdot \hat{r} )},
\end{equation}
where $t$ and $t'$ represent the time in the rest frame and moving frame, respectively. 
It is important to note that both the SSB frame and the source frame in Fig.~\ref{fig:frame} are rest frames. 
Specifically, the source frame corresponds to the rest frame during the early inspiral phase, when the kick velocity is negligible. 
The reference time point for the rest source frame is $t = -2000M$, where $M$ is the total mass and serves as a free scale. 
The transformation of kick velocity components between the SSB and source frames is detailed in Ref.~\cite{paper_Cao}.

\begin{figure}[ht]
    \begin{minipage}{\columnwidth}
        \centering
        \includegraphics[width=0.92\textwidth,
        trim=0 0 0 0,clip]{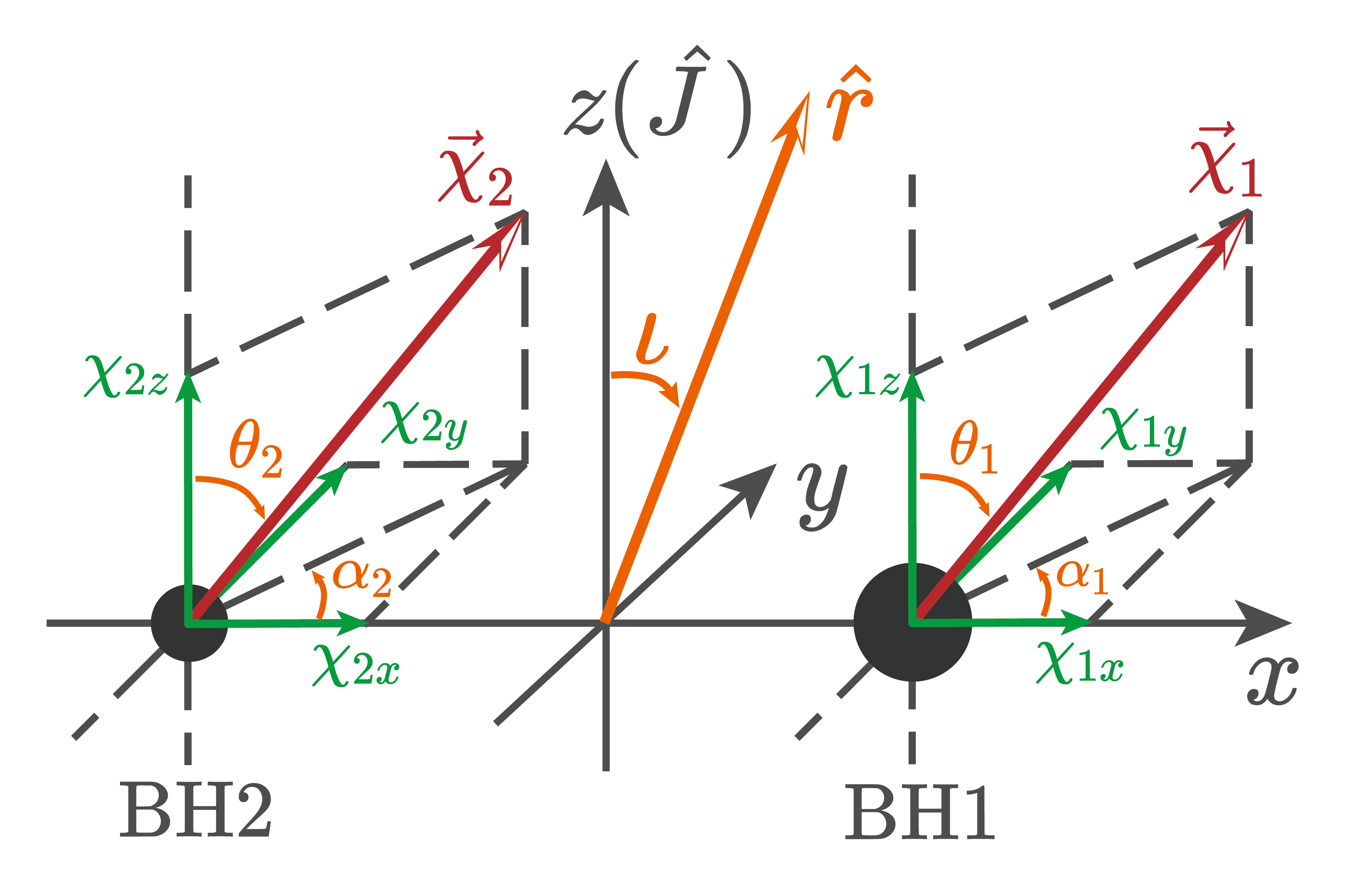}
        \caption{The diagram of the BBH spin components in the source frame. The BH with the larger mass is labeled as spin $\vec{\chi_1}$, and the BH with the smaller mass is labeled as spin $\vec{\chi_2}$.}\label{fig:spin}
    \end{minipage}
\end{figure}

\begin{figure}[ht]
    \begin{minipage}{\columnwidth}
        \centering
        \includegraphics[width=0.92\textwidth,
        trim=0 0 0 0,clip]{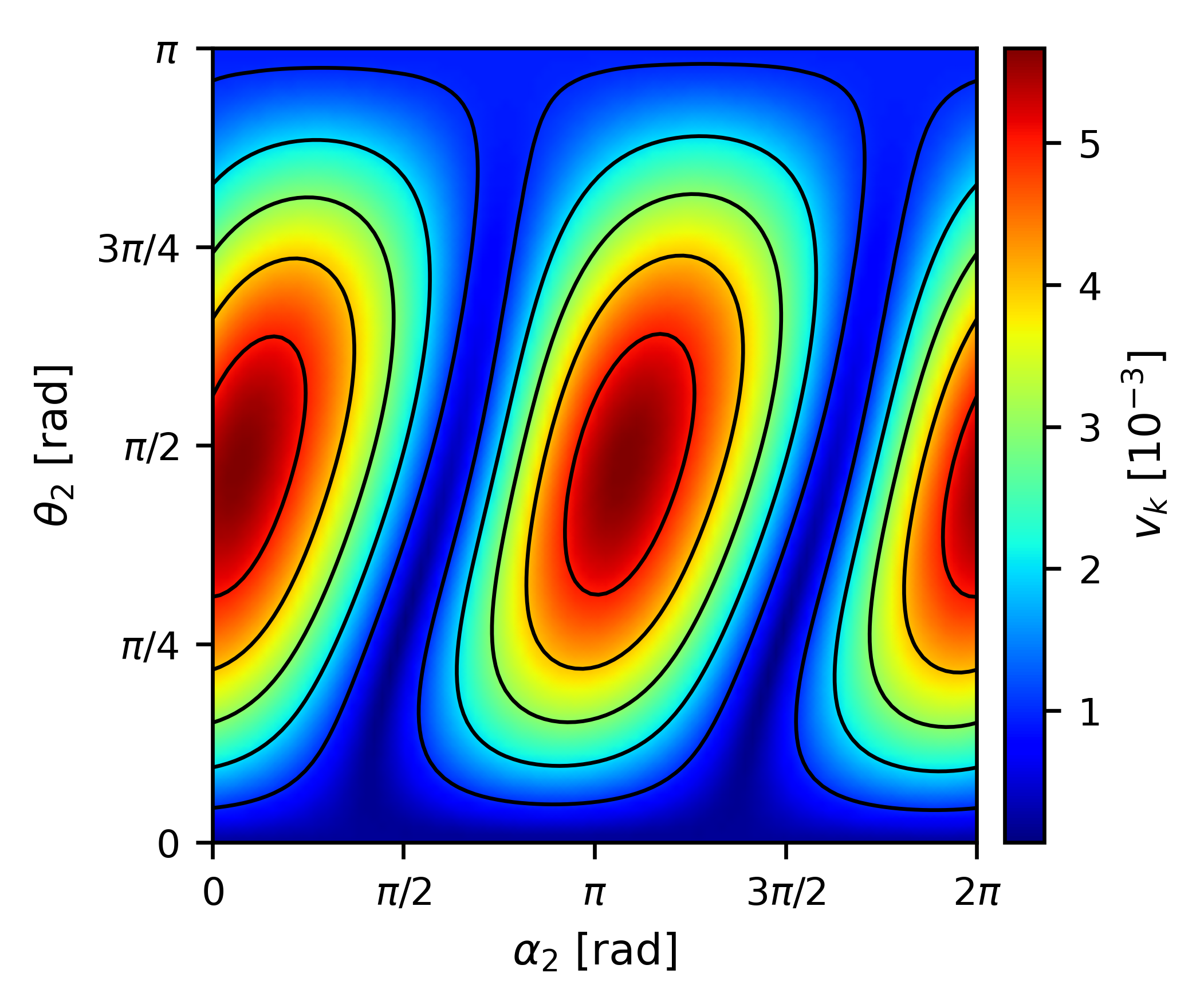}
        \caption{The final kick velocities for different spin angles. For an equal-mass BBH system, we set $\chi_1 = 0.4$ and $\chi_2 = 0.8$, with angles $\alpha_1 = \theta_1 = 0$, indicating that the spin of BH1 is entirely perpendicular to the orbital plane. All results are calculated using \texttt{SURRKICK} code~\cite{surrkick}.}\label{fig:spin_vk}
    \end{minipage}
\end{figure}

\begin{figure}[ht]
    \begin{minipage}{\columnwidth}
        \centering
        \includegraphics[width=0.9\textwidth,
        trim=0 0 0 0,clip]{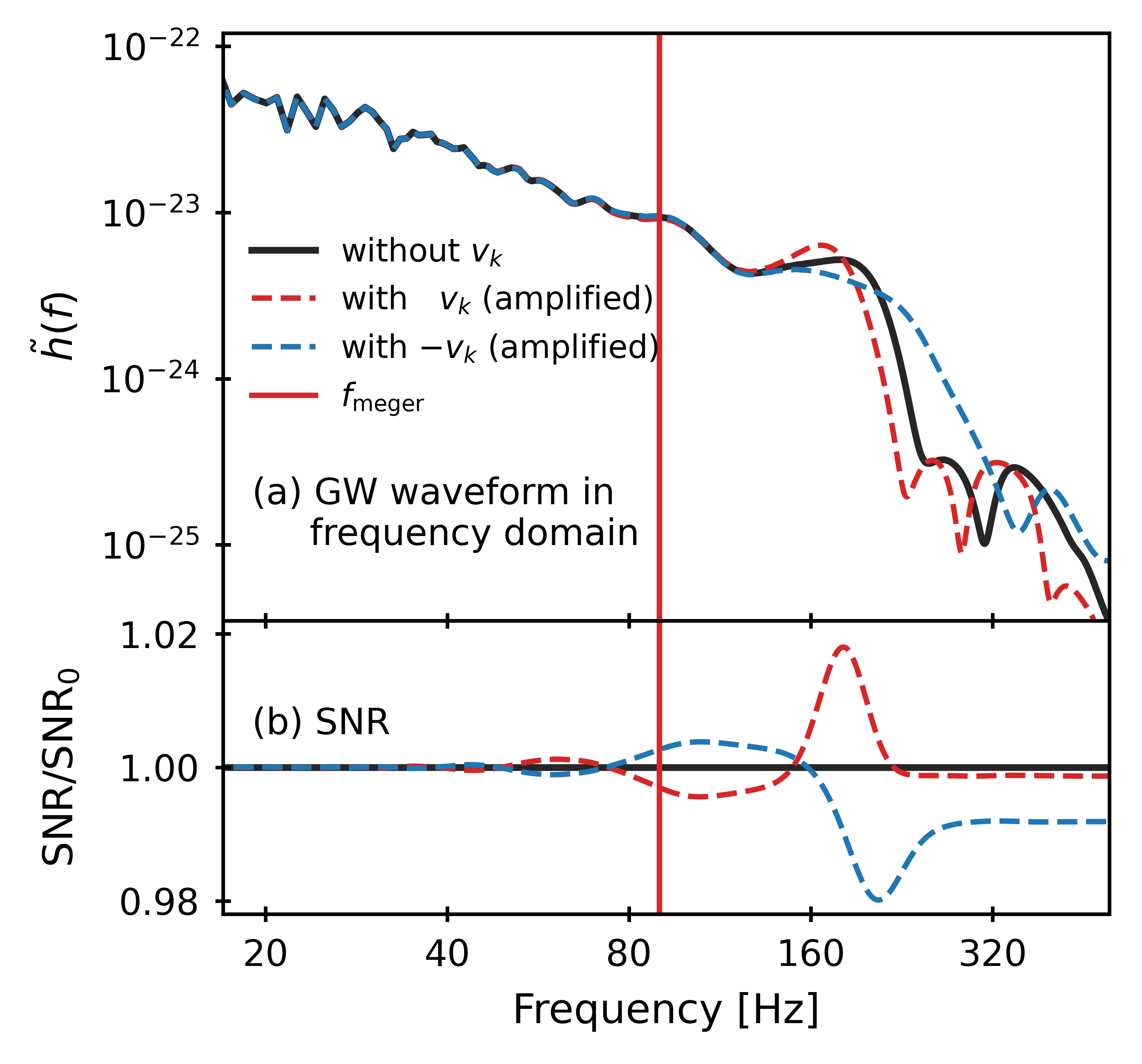}
        \caption{Comparison of frequency-domain GW waveforms with and without kick velocity. Note that the differences in the waveforms have been amplified, with the calculated kick velocity being 60 times the actual kick velocity. (a) illustrates the frequency-domain GW waveforms for the three cases (BBH parameters are shown in Fig.~\ref{fig:signal_time}). (b) provides a cumulative comparison of SNR over time, with the relative SNR values calculated using the case without kick velocity ($\mathrm{SNR_0}$) as the baseline. Calculation of SNR uses the LIGO design PSD (T1800044)~\cite{aLIGO}.}\label{fig:signal_frequency}
    \end{minipage}
\end{figure}

The kick velocity primarily depends on the system configuration (mass ratio and spins) and the dynamics of the merger, but not on the total mass (free scale)~\cite{kick2}. 
The definition of the BBH spins and their components is illustrated in Fig.~\ref{fig:spin}. 
The kick velocity for generic BBH systems can be accurately determined using a numerical-relativity surrogate model~\cite{kick_from_waveform,kick5}. 
We employ the Python program \texttt{SURRKICK} to calculate BBH kicks using the SXS surrogate model \texttt{NRSur7dq2}~\cite{surrkick}. 
The resulting kick velocities are used to transform GW waveforms from the rest frame to the moving frame.
Additionally, the components of the kick velocity are evaluated in the source's rest frame. 
To investigate the influence of spin states on the final kick velocity, we compute these effects, as shown in Fig.~\ref{fig:spin_vk}.

By varying the spin angles of BH2, we observe the impact of spin states on the final kick velocity. 
As illustrated in Fig.~\ref{fig:spin_vk}, higher kick velocities occur when the two spin directions are nearly perpendicular. 
Furthermore, degeneracy exists among various angles, meaning that different spin angles can yield the same final kick velocity. 
Note that we only present the final kick velocity; temporal variations in the kick velocity components may differ. 
For a more detailed analysis of the effects of mass ratios and complex spin states on kick velocity, refer to Ref.~\cite{surrkick}.

\begin{figure*}[ht]
    \begin{minipage}{\textwidth}
        \centering
        \includegraphics[width=0.95\textwidth,
        trim=0 5 0 0,clip]{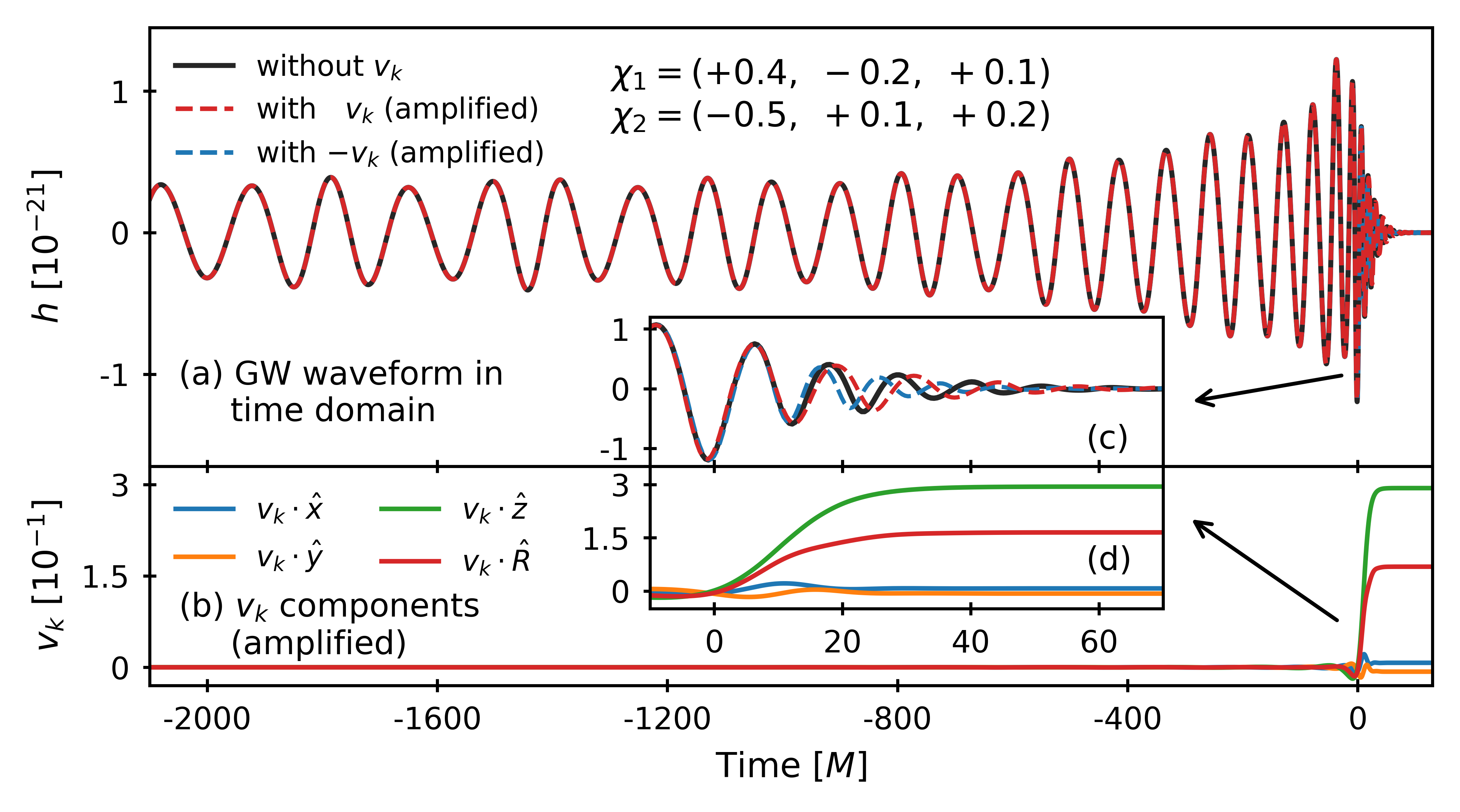}
        \caption{Comparison of time-domain GW waveforms with and without kick velocity. The GW waveform of an SBBH ($m_1=50\ \mathrm{M_\odot }, m_2=40\ \mathrm{M_\odot } ,D_L=500\ \mathrm{Mpc },e=0.1,\iota=\Theta=\Phi=\Psi=1,Mf_{\mathrm{ref} }=0.0066$) is observed by LIGO over 2000$M$ before coalescence. The time $t=0$ is the coalescence time. To emphasize the impact of kick velocity, all content in the image is based on an amplified kick velocity, with a magnification factor of 60. (a) represents the complete GW waveform, incorporating amplified kick velocities of 0, $v_k$, and -$v_k$, respectively. (b) illustrates the temporal evolution of the kick velocity components without amplification. The unit vector $\hat{R}$ denotes the direction of the line of sight from observer to source and $\hat{x}-\hat{y}-\hat{z}$ are the basis of the source frame. (c) and (d) provide enlarged views of (a) and (b) during the merger and ringing-down phases, respectively.}\label{fig:signal_time}
    \end{minipage}
\end{figure*}

Using \texttt{SEOBNRE} for GW waveform generation in the rest frame, \texttt{SURRKICK} for calculating time-dependent kick velocity components, and combining Eqs.~(\ref{eq:Lorentz1}) and (\ref{eq:Lorentz2}) for the Lorentz transformation, we construct GW waveforms that incorporate kick velocity effects for any BBH configuration. 
We select a specific SBBH and compute its waveform characteristics in both the time and frequency domains, considering scenarios with and without kick velocity. 
The results are illustrated in Figs.~\ref{fig:signal_frequency} and \ref{fig:signal_time}. 
To illustrate the differences with and without kick velocity, the kick velocities in Figs.~\ref{fig:signal_frequency} and \ref{fig:signal_time} have been artificially amplified.

From Fig.~\ref{fig:signal_time}, the kick velocity during the inspiral phase is relatively small. 
During the merger phase, the energy radiated by the BBH system leads to a significant increase in the kick velocity. 
In the ringdown phase, the remnant BH continues to move at a constant velocity. 
Consequently, differences in GW waveforms are most pronounced during the merger and ringdown phases. 
Specifically, the positive and negative projections of the kick velocity along the line-of-sight direction $\hat{R}$ induce redshift and blueshift effects in the GW waveform.
These effects are clearly visible in both the time and frequency domains, as shown in Figs.~\ref{fig:signal_frequency} and \ref{fig:signal_time}.

As shown in Fig.~\ref{fig:signal_frequency}, the kick velocity affects the GW waveform and alters the SNR. 
Although these changes are small, they may become observable for high-mass BBH systems and detectors with high sensitivity. 
Therefore, when observing high-SNR BBH events, the influence of kick velocity on GW waveforms must be considered. 
In Sec.~\ref{sec:Results}, we evaluate the impact of these differences across various detectors.

\section{Detectors}\label{sec:Detectors}
Several space-based detectors, including LISA, Taiji, and TianQin, are expected to launch in the mid-2030s~\cite{TianQin_white_paper,LISA_origin,ALIA}. 
These detectors will observe GWs in the millihertz frequency band, with their sensitivity and detection capabilities varying due to differences in arm length and orbital configuration. 
LISA and Taiji have arm lengths on the order of $10^6$ km, approximately one order of magnitude longer than TianQin. 
Additionally, while LISA and Taiji operate in heliocentric orbits, TianQin is positioned in a geocentric orbit, resulting in a distinct sensitive frequency range.

Each detector consists of a triangular formation of three spacecraft. 
Taiji proposes three orbital configurations: Taiji-p, Taiji-c, and Taiji-m, which maintain identical sensitivity levels but differ in their network configurations with LISA~\cite{Taiji1,Taiji2,Taiji3}. 
Similarly, TianQin's sensitivity varies based on the orientation of the constellation plane's normal direction~\cite{TianQin_configuration}. 
Detailed descriptions of these configurations can be found in Refs.~\cite{paper2,paper3,paper4}.

In contrast to the planned space-based detectors, only ground-based detectors are currently operational. 
These include the LVK network's four facilities: LIGO Hanford (H1), LIGO Livingston (L1), Virgo (V1), and KAGRA (K1)~\cite{LVK1,LVK2,LVK3}. 
These detectors employ L-shaped interferometer designs, while the third-generation detector, ET, utilizes a triangular configuration that effectively functions as a network of three independent detectors~\cite{ET1,ET2,ET3}.

Ground-based detectors are affected by various noise sources, such as quantum noise, seismic noise, gravity-gradient noise, and thermal noise. 
For our calculations, we use the design noise power spectral density (PSD)~\cite{pycbc1,pycbc2,pycbc3,pycbc4}.
For space-based detectors, we account for the impact of time-delay interference (TDI) on both signal and noise, focusing on the PSD of secondary noise components: displacement noise and acceleration noise. 
These are calculated using the XYZ channels of second-generation TDI technology (see Ref.~\cite{paper1}).
The noise PSD of these detectors is used to compute the inner product, which is essential for calculating the SNR and error in Secs.~\ref{sec:Methodology} and \ref{sec:Results}.

\section{Methodology}\label{sec:Methodology}
\subsection{Data analysis}
In the field of GW data processing and analysis, the inner product is conventionally defined as~\cite{SNR}
\begin{equation}\label{eq:inner_product}
    (a|b)=4\text{Re}\left[\int_0^{\infty}\frac{\tilde{a}^*(f)\tilde{b}(f)}{S_n(f)}\mathrm{d}f\right],
\end{equation}
where $S_n(f)$ is the noise PSD of detector, $\tilde{a}(f)$ and $\tilde{b}(f)$ are the Fourier transforms of $a(t)$ and $b(t)$, respectively.
For a time-domain signal $h(t)$, the SNR can typically be expressed as~\cite{SNR} 
\begin{equation}
    \mathrm{SNR}^2 =(h|h) .
\end{equation}
For a detector network, the joint SNR can be expressed as the sum of individual inner products~\cite{paper4}:
\begin{equation}\label{eq:sum_inner_products}
    \mathrm{SNR}^2=\sum_i{\mathrm{SNR}_i^2}=\sum_i{(h_i|h_i)} .
\end{equation}
The SNR is a critical metric for assessing the detectability of GW signals. 
In this study, we adopt an SNR threshold of 8: signals with $\mathrm{SNR} \geq 8$ are considered \textit{detectable}, while those with $\mathrm{SNR} < 8$ are deemed \textit{undetectable}~\cite{SNR_threshold1,SNR_threshold2}.

In addition to SNR, the accuracy requirement of waveform modeling is essential for extracting scientific information efficiently. 
Utilizing an appropriate precision waveform ensures the comprehensive extraction of scientific information from the data while avoiding unnecessary computational resource burdens caused by excessive accuracy requirements~\cite{paper3}.
We employ the method from Ref.~\cite{Accuracy_requirement} to compare the rest-frame waveform $h_0$ with the waveform $h_k$, which includes the kick velocity effect. 
The discrepancy between these waveforms is quantified by the Error:
\begin{equation}
    \mathrm{Error}=(\delta h|\delta h),
\end{equation}
where $\delta h = h_0 - h_k$ denotes the waveform error.
For a detector network, the joint Error is calculated similarly to Eq.~(\ref{eq:sum_inner_products}).

This method employs a imple yet strict accuracy requirement calculation method. 
The $\delta h$ is assessed using the Fisher Information Matrix. 
If $\mathrm{Error} < 1$, the waveforms are considered \textit{indistinguishable}, whereas if $\mathrm{Error} \geq 1$, they are \textit{distinguishable}.
In practical terms, $\mathrm{Error} < 1$ implies that the detector cannot differentiate between the waveforms, and higher accuracy is unnecessary. 
When $\mathrm{Error} \geq 1$, less accurate waveforms may compromise measurements, necessitating consideration of the kick velocity effect. 
Compared with mismatch, the method used to compute Error imposes a more stringent requirement, as detailed in Refs.~\cite{Accuracy_requirement, Accuracy_more1, Accuracy_more2, paper3}. 
Additionally, the calculation of Error requires only a single inner product computation, making it more efficient than the three inner product calculations required for mismatch. 
This efficiency is a key reason for our choice to use Error as the metric for waveform accuracy in this study. 
Consequently, we rely on SNR and Error as the two primary metrics to comprehensively assess the impact of kick velocity on waveforms and to evaluate the performance of various detectors.
\subsection{BBH source selection}
In this paper, we consider two scenarios for BBH source selection. 
First, to study the impact of kick velocity under varying parameters, we randomly sample values from different parameter spaces and evaluate the results across multiple detectors. 
Second, to reflect practical detection capabilities, we select representative BBH population models to determine whether detectors can discern the kick velocity effect.

\begin{table}[ht]
    \centering
    \renewcommand{\arraystretch}{1.5}
    \caption{Parameter distribution used in calculation. $U[a,b]$ represents a uniform distribution from $a$ to $b$.}\label{tab:parameters}
    \begin{tabular*}{\columnwidth}{@{\extracolsep{\fill}}lrr@{}}
    \hline
     Parameter & Distribution & Unit\\
    \hline
    log($M$) [space] & $U[4,8]$ & $\mathrm{M_\odot } $\\
    $M$ [ground] & $U[10,180]$ & $\mathrm{M_\odot } $\\
    $q$ & $U[1,2]$& \\
    log($D_L$) [space] & $U[0.25,2.5]$ & Gpc\\
    log($D_L$) [ground] & $U[1.5,4]$ & Mpc\\
    log($e$) & $U[-4,-1]$& \\
    $t_c$ & $U[0,1]$ & yrs\\
    $\chi_{1/2}$ & $U[0,0.8]$ & \\
    $\phi_0,\Phi,\Psi,\alpha_{1/2}$ & $U[0,2\pi]$ & rad\\
    $\Theta,\theta_{1/2}$ & $\arcsin(U[-1,1])$ & rad\\
    $\iota$ & $\arccos(U[-1,1])$ & rad\\
    \hline
    \end{tabular*}
\end{table}

In the generation of GW signals, we take into account a total of 16 parameters, listed as follows:
\begin{equation}
    \begin{aligned}
    \boldsymbol{\xi}=\{M,q,D_L,e,\phi_0,\iota,t_c, \Theta,\Phi,\Psi\\
        \chi_1,\alpha _1,\theta_1,\chi_2,\alpha _2,\theta_2\},
    \end{aligned}
\end{equation}
where $q$ is the mass ratio, $D_L$ is the luminosity distance, $e$ is the orbital eccentricity, $\phi_0$ is the initial phase, $\iota$ is inclination angle, $t_c$ is the coalescence time, $(\Theta,\Phi)$ is the sky position, $\Psi$ is polarization angle, and $(\chi_1,\alpha _1,\theta_1,\chi_2,\alpha _2,\theta_2)$ are the spin parameters (see Fig.~\ref{fig:spin}).
We set the reference frequency $f_{\mathrm{ref} }$ for defining eccentricity and spin to $0.95f_0$, where $f_0$ represents the GW frequency at $t=-2000M$.
The parameter ranges, detailed in Table~\ref{tab:parameters}, are based on Refs.~\cite{LISA_population1,LISA_population2,paper2,paper3} and \texttt{SURRKICK}~\cite{surrkick}.
We employ a random sampling method to select the parameters of the BBH source based on this distribution, ensuring that sources with diverse parameters are as comprehensively covered as possible within the parameter space. 
For cosmological modeling, we adopt the $\mathrm{\Lambda CDM}$ model with parameters from the \textit{Planck} 2018 results~\cite{Planck_2018}: Hubble constant $H_0 = 67.37\ \mathrm{km}\ \mathrm{s}^{-1} \mathrm{Mpc}^{-1}$, matter density parameter $\Omega_m = 0.315$, and dark energy density parameter $\Omega_\Lambda = 0.685$. 
Note that the parameters in Table~\ref{tab:parameters} are in the source frame; redshift effects are accounted for by scaling the mass by $(1+z)$.

In addition to employing random sampling in the parameter space to assess waveform modeling accuracy in Sec.~\ref{subsec:effect_of_kick_velocity}, we also consider several representative BBH population models in Sec.~\ref{subsec:effect_on_detection} to evaluate the impact of kick velocity in actual observations.

For the SBBH observed by ground-based detector, there is minimal variation between different models after applying LVK observational constraints~\cite{LIGO_population1,LIGO_population2,LIGO_population3}. 
We select the power-law distribution for the mass model and adopt the Madau-Dickinson star formation rate~\cite{Star_Formation}. 
On this basis, two types of models are defined: \texttt{popA}, which does not consider the peak component of the mass function, and \texttt{popB}, which incorporates this peak component. 
For the MBHB observed by space-based detector, astrophysics proposes various population models in the absence of actual GW event constraints. 
Specifically, we consider three primary population models: \texttt{pop3}, \texttt{Q3nd}, and \texttt{Q3d}~\cite{TianQin_MBHB}. 
The \texttt{pop3} model represents a light seed model where MBHB seeds originate from the remnants of PopIII stars~\cite{pop3}. 
The \texttt{Q3} models represent heavy seed model where MBHB seeds originate from the collapse of protogalactic disks~\cite{Q3}. 
Furthermore, \texttt{Q3d} incorporates the delay between MBHB and galaxy mergers, whereas \texttt{Q3nd} does not account for this delay. 
For the above five BBH population models, we utilize \texttt{GWToolbox} to generate source parameters for calculation in Sec.~\ref{sec:Results}~\cite{GWToolbox1,GWToolbox2}.

\section{Results}\label{sec:Results}
\subsection{effect of kick velocity}\label{subsec:effect_of_kick_velocity}
\begin{figure}[ht]
    \begin{minipage}{\columnwidth}
        \centering
        \includegraphics[width=0.93\textwidth,
        trim=0 0 0 0,clip]{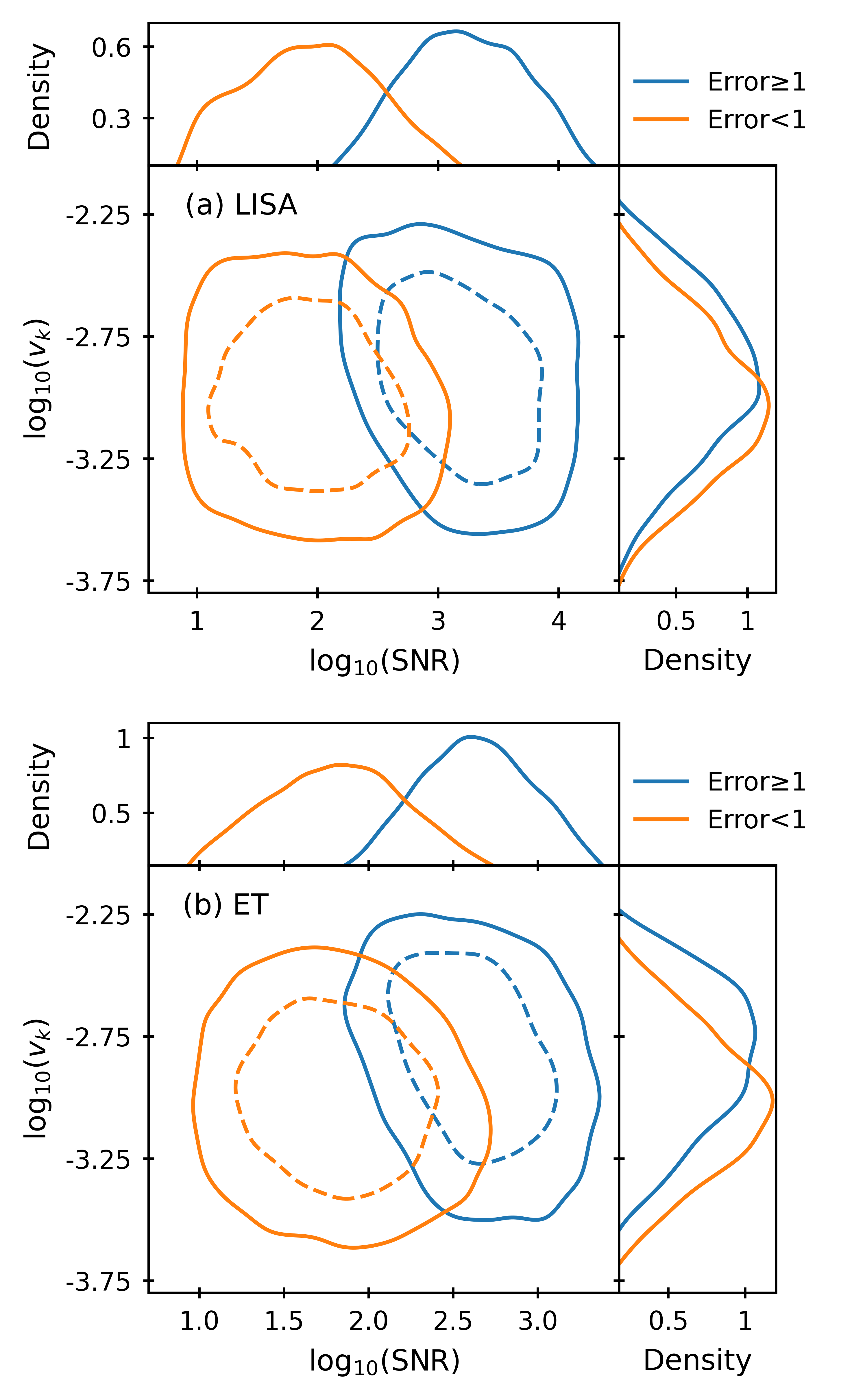}
        \caption{Comparison and distribution of SNR and kick velocity. (a) and (b) show the results for LISA and ET, respectively. The solid and dashed lines represent the 90\% and 50\% confidence intervals. The smaller plots on the right and top illustrate the probability densities of kick velocity and SNR.}\label{fig:SNR_vk}
    \end{minipage}
\end{figure}

We construct two datasets, each consisting of 10,000 SBBHs and MBHBs, as detailed in Table~\ref{tab:parameters}. 
By simulating GW signals from these datasets and analyzing them with various detectors, we compute the corresponding SNRs and Errors. 
All results presented correspond to detectable GW signals, i.e., SNR$\geq8$. 
To examine the impact of kick velocity, we present the results for LISA and ET in Fig.~\ref{fig:SNR_vk}.

From Fig.~\ref{fig:SNR_vk}, it can be observed that when the SNR is low, the Error for distinguishing between $h_0$ and $h_k$ waveforms remains below 1, indicating that the detector cannot reliably differentiate between these waveforms. 
Conversely, with a sufficiently high SNR, the two waveforms become distinguishable regardless of the kick velocity. 
For intermediate SNR values, a smaller kick velocity requires a higher SNR to achieve waveform differentiation. 
The density distribution further shows that distinguishable cases generally exhibit significantly higher SNR and kick velocities compared to indistinguishable cases. 
Since increased kick velocity has a more pronounced effect on the waveform and a larger SNR enhances signal strength, the detector's ability to distinguish between waveforms improves. 
Thus, the influence of kick velocity on waveform characteristics must be carefully considered.

\begin{figure}[ht]
    \begin{minipage}{\columnwidth}
        \centering
        \includegraphics[width=0.93\textwidth,
        trim=0 0 0 0,clip]{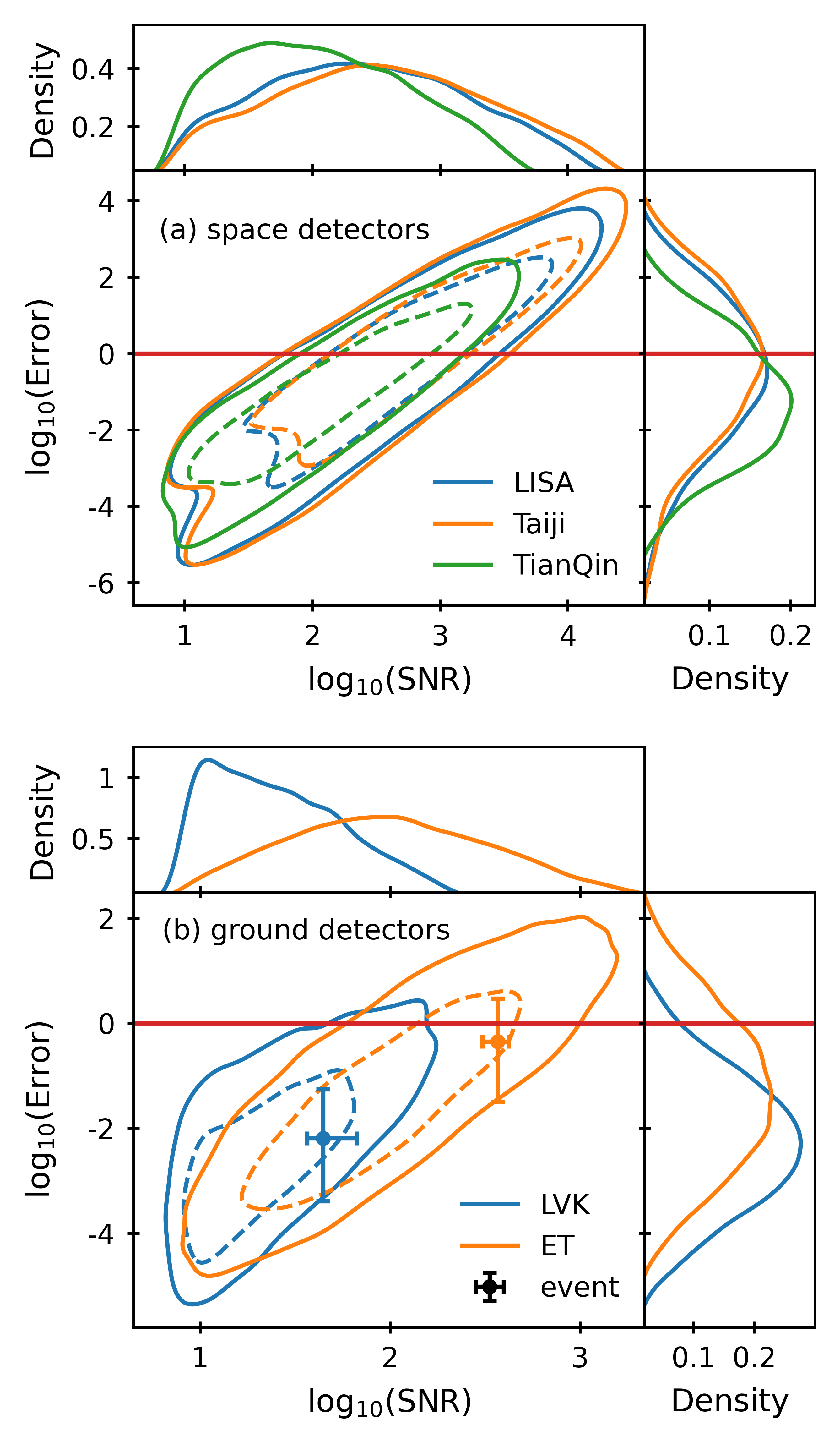}
        \caption{Comparison and distribution of SNR and Error. (a) and (b) depict the results obtained from space- and ground-based detectors, respectively. The lines are similar to those in Fig.~\ref{fig:SNR_vk}. The red horizontal line represents Error = 1, which is the dividing line between distinguishable and indistinguishable. The \textit{event} in (b) represents the result of GW150914-like events ($m_1=35.6_{-3.1}^{+4.7}\ \mathrm{M_\odot } ,m_2=30.6_{-4.4}^{+3}\ \mathrm{M_\odot } ,D_L=440_{-170}^{+150}\ \mathrm{Mpc }$~\cite{GWTC1}) in LVK and ET.}\label{fig:SNR_Error}
    \end{minipage}
\end{figure}

In this study, the kick velocity is derived from the parameters of the BBH system and is not treated as an independent parameter. 
Consequently, our primary focus is on the distribution of SNR and Error across various detectors. 
The results are presented in Fig.~\ref{fig:SNR_Error}.

As illustrated in Fig.~\ref{fig:SNR_Error}, there is a roughly positive correlation between Error and SNR. 
A higher SNR is typically associated with a larger Error, as inferred from the inner product calculation (Eq.~(\ref{eq:inner_product})). 
For space-based detectors, GW signals with an SNR below 100 are nearly indistinguishable, making it impossible to observe the effect of kick velocity. 
Conversely, almost all signals with SNRs exceeding 2500 can be clearly distinguished. 
Additionally, the density distribution reveals that Taiji performs best, exhibiting the highest SNR and Error values, followed closely by LISA, while TianQin shows a relatively larger gap. 
These outcomes are directly linked to the arm length and sensitivity of the detectors~\cite{paper1,paper2,paper3}. 
In terms of overall distribution, more than one-third of the cases are distinguishable.

For ground-based detectors, GW signals with an SNR below 70 are almost indistinguishable, making discernible signals rare for second-generation detectors. 
Due to the superior sensitivity of ET compared to LVK, some GW signals can still be distinguished. 
For instance, for GW150914-like events, the Error margins reported by the LVK are all below 1, rendering the effect of kick velocity unobservable. 
In contrast, the ET may potentially detect this effect. 
Additionally, limited by the BBH source characteristics and detector capabilities, the proportion of distinguishable cases in ground-based detectors is smaller than that in space-based detectors. 

In summary, within the defined parameter space, space-based detectors have the potential to observe the effect of kick velocity, which is nearly impossible for second-generation ground-based detectors but remains a possibility for third-generation ground-based detectors. 
In the following section, we explore the different population models of BBH and the detector network in detail.
\subsection{effect on detection}\label{subsec:effect_on_detection}
\begin{figure*}[ht]
    \begin{minipage}{\textwidth}
        \centering
        \includegraphics[width=0.95\textwidth,
        trim=0 5 0 0,clip]{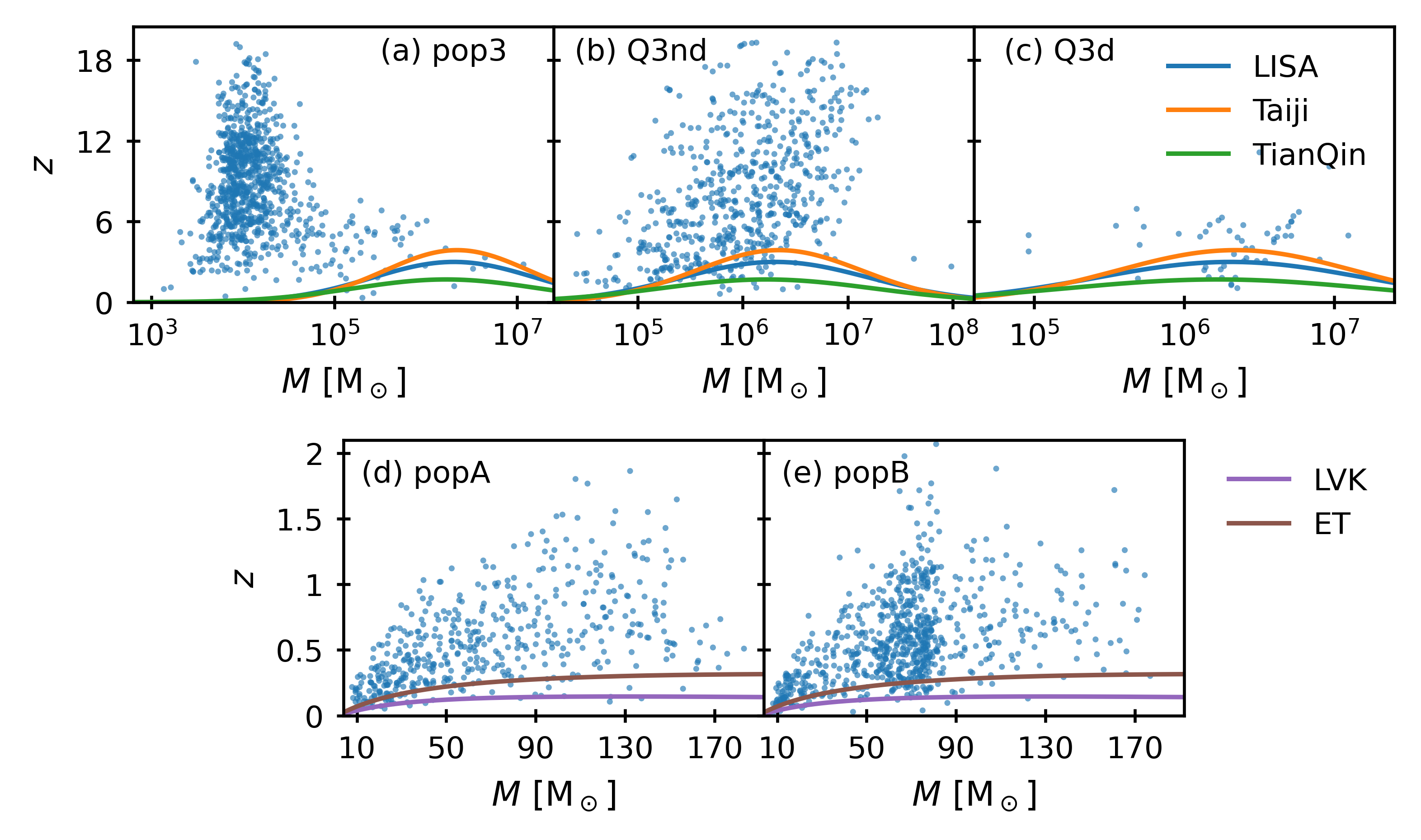}
        \caption{Comparison of different population models with space- and ground-based detectors. (a), (b), and (c) represent the three population models \texttt{pop3}, \texttt{Q3nd}, and \texttt{Q3d} of MBHB, respectively. Meanwhile, (d) and (e) correspond to the two population models \texttt{popA} and \texttt{popB} of SBBH. The lines in the figure are derived from polynomial fitting based on the condition Error$\approx$1, as described in Sec.~\ref{subsec:effect_of_kick_velocity}, serving a schematic purpose only. The region below the line indicates the discernible scenario, whereas the region above the line represents the indistinguishable scenario. The results in Tables~\ref{tab:population_space} and \ref{tab:population_ground} are obtained through calculations, not through the fitted lines in the figure.
        }\label{fig:population}
    \end{minipage}
\end{figure*}

The calculations within the parameter space strike a balance between waveform modeling accuracy and computational resource requirements. 
By utilizing diverse population models, we gain a deeper understanding of potential scenarios encountered in actual observations. 
Specifically, Table~\ref{tab:population_space} presents the results from three MBHB population models observed by the space-based detectors, while Table~\ref{tab:population_ground} details findings from two SBBH population models observed by ground-based detectors.
Additionally, Figure~\ref{fig:population} illustrates the distribution of total mass $M$ and redshift $z$ across different population models.
Notably, the fitting lines in Fig.~\ref{fig:population} are included solely for illustrative purposes and are not employed in the calculations.

\begin{table}[ht]
    \centering
    \renewcommand{\arraystretch}{1.5}
    \caption{Distinguishable proportions of various space-based detectors and networks within three MBHB population models. Herein, \textit{TJ} denotes Taiji, while \textit{TQ} refers to TianQin.}\label{tab:population_space}
    \begin{tabular*}{\columnwidth}{@{\extracolsep{\fill}}lrrr@{}}
    \hline
        & \texttt{pop3} & \texttt{Q3nd} & \texttt{Q3d}\\
    \hline
    LISA  & 1.69 \% & 23.49 \% & 40.0 \%\\
    Taiji  & 3.15 \% & 33.02 \% & 54.0 \%\\
    TianQin  & 0.79 \% & 7.78 \% & 14.0 \%\\
    LISA+TJp  & 3.6 \% & 39.05 \% & 62.0 \%\\
    LISA+TJc  & 3.37 \% & 38.73 \% & 62.0 \%\\
    LISA+TJm  & 3.93 \% & 39.68 \% & 62.0 \%\\
    LISA+TQ  & 1.91 \% & 24.92 \% & 46.0 \%\\
    TJ+TQ  & 3.15 \% & 33.49 \% & 56.0 \%\\
    \hline
    \end{tabular*}
\end{table}

\begin{table}[ht]
    \centering
    \renewcommand{\arraystretch}{1.5}
    \caption{Distinguishable proportions of various ground-based detectors and networks within three MBHB population models.}\label{tab:population_ground}
    \begin{tabular*}{\columnwidth}{@{\extracolsep{\fill}}lrr@{}}
    \hline
    & \texttt{popA} & \texttt{popB}\\
    \hline
    L1  & 0.21 \% & 0.39 \%\\
    LVK  & 0.91 \% & 1.03 \%\\
    ET  & 12.76 \% & 16.32 \%\\
    LVK+ET  & 12.96 \% & 16.84 \%\\
    \hline
    \end{tabular*}
\end{table}

The parameter distribution of BBH sources varies significantly across different population models. 
As illustrated in Fig.~\ref{fig:population}, the three MBHB population models exhibit notable differences. 
The number of MBHBs varies across different population models.
In our simulation, the numbers of MBHBs used for the three models (\texttt{pop3}, \texttt{Q3nd}, and \texttt{Q3d}) are 890, 630, and 50, respectively.
Specifically, the \texttt{pop3} model features a smaller MBHB source mass, primarily concentrated within the range of $10^3$ to $10^5$ $\mathrm{M_\odot}$, placing most of these sources outside the detector's resolution range. 
In contrast, the mass ranges for the \texttt{Q3d} and \texttt{Q3nd} models are similar, predominantly falling between $10^5$ and $10^7$ $\mathrm{M_\odot}$. 
While the \texttt{Q3nd} model exhibits a higher redshift distribution, the MBHB sources of the \texttt{Q3d} model have a redshift less than 8, with the lowest total number among the three models. 
Consequently, the distinguishable proportions in Table~\ref{tab:population_space} are highest for the \texttt{Q3d} model, followed by the \texttt{Q3nd} model, and lowest for the \texttt{pop3} model.

In contrast to the MBHB population model, the two SBBH population models exhibit similar characteristics. 
Considering the large number of SBBHs in the population model, we set the number of SBBHs to 1000 for both of these population models in our simulation.
Compared to the \texttt{popA} model, the \texttt{popB} model includes an additional mass peak component, which results in a marginally higher distinguishable proportion. 
Overall, the outcomes of the \texttt{popA} and \texttt{popB} models are largely consistent, as illustrated in Table~\ref{tab:population_ground}.

For a single space-based detector, the conclusions presented in Table~\ref{tab:population_space} are consistent with those illustrated in Fig.~\ref{fig:SNR_Error}. 
When considering detector networks, the results remain largely similar to those obtained from single detectors.  
The optimal configuration is LISA+TJ, as these represent the two most effective individual detectors. 
TJ+TQ ranks second, while LISA+TQ exhibits the lowest percentage of resolvable signals. 
Unlike in single-detector scenarios, variations in Taiji configurations do not significantly impact the overall results. 
However, within the LISA+TJ network, different constellation configurations yield slightly varying outcomes. 
Specifically, the LISA+TJm constellation plane has the largest angle in the normal direction, providing the highest resolution capability and thus achieving the best performance. 
Conversely, the LISA+TJc constellation plane, sharing an identical normal orientation, is the least distinguishable among the three configurations (for further analysis, see Ref.~\cite{paper2}).

Unlike space-based detectors, the results obtained from ground-based detectors are significantly more succinct. 
The performance gap between second-generation and third-generation ground-based detectors is substantial. For instance, the current network of four detectors, collectively referred to as LVK, can resolve signals with a probability of less than 1\%, whereas ET achieves distinguishable proportions exceeding 10\%. 
When considering the combined capabilities of LVK+ET, the improvement remains modest compared to ET alone. 
LIGO has limited capability to observe the effects of kick velocity, whereas ET demonstrates potential in identifying some of these phenomena.

In future observations, it is necessary to account for kick velocity. 
According to the optimistic model, over half of the MBHB sources are expected to exhibit observable kick effects in space-based detectors. 
Additionally, third-generation ground-based detectors are anticipated to detect kick effects from more than 10\% of SBBH sources. 
Consequently, employing more precise waveform modeling will significantly enhance signal processing and the extraction of scientific information from BBH sources.

\section{Conclusion}\label{sec:Conclusion}
In this paper, we investigate the influence of kick velocity on GW waveforms and its implications for detection by both space- and ground-based detectors. 
By applying the Lorentz tensor transformation, we analyze the GW waveform between moving and rest frames, utilizing \texttt{SURRKICK} to generate time-dependent kick velocity components and \texttt{SEOBNRE} to produce the corresponding GW waveforms. 
For space-based detectors, we consider LISA, Taiji, and TianQin for observing MBHBs; for ground-based detectors, we focus on LVK and ET. 
We sample BBH source data within the parameter space and calculate the SNR and Error for each detector. 
Additionally, we employ several representative BBH population models to simulate potential future detection scenarios. 
Through these methodologies, we systematically evaluate and analyze the impact of kick velocity on GW observations.

Our results emphasize the critical balance between achieving accurate waveform modeling under the influence of kick velocity and managing computational resources. 
For waveform modeling in parameter space (Tabel~\ref{tab:parameters}), GW signals with higher SNR are more likely to exhibit observable effects from kick velocity (see Fig.~\ref{fig:SNR_Error}). 
Specifically, within the operational parameters of space-based detectors, nearly 50\% of GW signals require consideration of kick velocity effects on their waveforms. 
For ground-based detectors, this proportion is less than one-third for ET observations. 
Overall, our findings underscore the necessity of incorporating kick velocity effects in waveform modeling for both space- and ground-based detectors.

When evaluating observations of several typical BBH population models (see Fig.~\ref{fig:population} and Tables~\ref{tab:population_space}-\ref{tab:population_ground}), we find that in an optimistic scenario, space-based detector networks can observe the kick effect in over 60\% of MBHB sources, while in a pessimistic scenario, this proportion drops to 3$\sim$4\%. 
Second-generation ground-based detectors are expected to observe the kick effect in approximately 1\% of SBBH sources, whereas third-generation detectors may observe this effect in up to 16\% of cases. 
Overall, space-based detectors have a significantly higher possibility of detecting the effects induced by kick velocities. 
This capability is anticipated to yield valuable scientific insights, facilitating further research into the structure and evolution of galaxies.

In future research, we plan to significantly broaden our scope. 
Specifically, we intend to extend our investigations to modified gravity theories, incorporating GW waveforms with six polarization modes. 
Additionally, we will explore the practical applications of GW waveforms that account for kick velocities in the real data processing~\cite{kick_GWTC}. 
We can employ kick inference derived from numerical relativity fitting to estimate the retention probability within clusters and investigate the hierarchical mergers of BHs~\cite{hierarchical_mergers1,hierarchical_mergers2}.
This will involve using matched filtering techniques to perform Markov Chain Monte Carlo inference on existing LVK event data. 
We anticipate that these efforts will yield comprehensive and valuable insights into the effects of kick velocities, providing a robust foundation for further research in this field.

\begin{acknowledgements}
This work was supported by the National Key Research and Development Program of China (Grant No. 2023YFC2206702), the Fundamental Research Funds for the Central Universities Project (Grant No. 2024IAIS-ZD009), the National Natural Science Foundation of China (Grant No. 12347101), and the Natural Science Foundation of Chongqing (Grant No. CSTB2023NSCQ-MSX0103). 
\end{acknowledgements}

\bibliography{references}
\end{document}